\documentclass{article} %%% use \documentstyle for old LaTeX compilers
\usepackage{arxiv}

\usepackage[english]{babel} %%% 'french', 'german', 'spanish', 'danish', etc.
\usepackage{amssymb}
\usepackage{amsmath}
\usepackage{txfonts}
\usepackage{mathdots}
\usepackage[classicReIm]{kpfonts}
\usepackage[dvips]{graphicx} %%% use 'pdftex' instead of 'dvips' for PDF output

\usepackage{subfigure}

\usepackage[utf8]{inputenc}
\usepackage{algorithm} %ctan.org\pkg\algorithms
\usepackage{algpseudocode}
\usepackage{tikz}
\usetikzlibrary{backgrounds,fit,decorations.pathreplacing, shapes}  % TikZ libraries
\usepackage{tkz-graph}

\usepackage[super]{nth}
\usepackage{adjustbox}
\usepackage{nameref}

\tikzstyle{vertex}=[circle,draw=black, fill=white,sloped,minimum size=17pt,inner sep=5pt]

\usepackage{tikz}
\usetikzlibrary{decorations.pathreplacing,calc}

\usepackage{enumerate}
\usepackage{amsmath}
\usepackage{graphicx}
\usepackage{wrapfig}
\usepackage{lscape}
\usepackage{rotating}
\usepackage{epstopdf}

\usepackage{caption}
\usepackage{url}
\usepackage{hyperref}
\usepackage{float}
\usepackage[super]{nth}
\usepackage[labelfont=bf]{caption}
\usepackage[labelsep=period]{caption}
\usepackage{pgfplots} % LaTeX

\pgfplotsset{compat = newest}

\newcommand{\figref}[1]%
{Figure \ref{#1}%
}

\newcommand{\tableref}[1]%
{Table \ref{#1}%
}

\newcommand{\algorithmref}[1]%
{Algorithm \ref{#1}%
}

\newcommand{\sectionref}[1]%
{Section \ref{#1}%
}

\newcommand{\lineref}[1]%
{Line \ref{#1}%
}

\algnewcommand{\LineComment}[1]{\State \(\triangleright\) #1}

\usepackage{newunicodechar}
\newunicodechar{−}{-}

\title{A System for Efficient Communication between Patients and Pharmacies}

\author{Rui Portocarrero Sarmento \\
LIAAD-INESC TEC \\
PRODEI - Faculty of Engineering, University of Porto \\
mail@ruisarmento.com \And
André Tarrinho \\
Lloyds Pharmacy \\
andretarrinho@gmail.com \And
Pedro Câmara \\
FEUP, University of Porto \\
up201304073@fe.up.pt \And
Vera Costa \\
FEUP, University of Porto \\
veracosta@fe.up.pt}

\begin{document}

\long\def\/*#1*/{}

\maketitle

\begin{abstract}
When studying human-technology interaction systems, researchers thrive to achieve intuitiveness and facilitate the people's life through a thoughtful and in-depth study of several components of the application system that supports some particular business communication with customers. Particularly in the healthcare field, some requirements such as clarity, transparency, efficiency, and speed in transmitting information to patients and or healthcare professionals might mean an important increase in the well-being of the patient and productivity of the healthcare professional. In this work, the authors study the difficulties patients frequently have when communicating with pharmacists. In addition to a statistical study of a survey conducted with more than two hundred frequent pharmacy customers, we propose an IT solution for better communication between patients and pharmacists.
\end{abstract}


\section{Introduction}

 Nowadays, for both companies and persons, time management is something crucial. Community pharmacies are an important pillar of the national health system and are a privileged point of contact with patients. However, this contact is sometimes subject to long queues, especially during periods of minimum service or weekends. Patients waste a lot of time looking for the medication they need.

 The main questions and issues patients had when in need of using pharmacy retail stores were previously addressed by several apps and websites that support the search for nearby pharmacy stores, inform the patient if they are open and provide general contacts like telephone numbers or email. Although this seems to be a standard method for obtaining the medicines in a patient´s prescription, at our best knowledge there is no previous inspection of the public opinion about their needs and if a better option for improving the efficiency of such means of communication between healthcare pharmacists and patients would exist. Therefore, we posed some questions about these issues to ourselves and to some patients. It was not clear whether these were present in the minds of the majority of the patients, exactly because the systems seem to be established and resilient. We will show that there are possible improvements that the previous assumption might not be quite right. The main questions are:

\begin{enumerate}
\item  Would it be possible for a mean of communication between a patient and a pharmacy which does not involve physical displacement or telephone contact that often disturbs the normal operation of the pharmacy?
\item  Would it be possible for a means of communication to be made available to the user to check whether the pharmacies in the area had the necessary prescription medicines for themselves, namely low-volume or out-of-store drugs?
\end{enumerate}

 With these issues in mind, the authors did a survey that had hundreds of respondents. The results were assessed with statistical methods to provide an answer to both questions. Additionally, in this document, the authors present MEDLOC, an app created to achieve a better solution for medicine search by patients and a more efficient response from the pharmacists. MEDLOC would allow smartphone and tablet users to have information on the availability of medications at nearby pharmacies, and would allow the user to better manage his prescriptions.

\section{Background }

 Community pharmacists are major contributors to the health care system around the world. The role of the pharmacist has evolved throughout the years, shifting from just selling medicines to more a patient-centered approach with an increased offer of health services \cite{Viswanathan2015}. Pharmacists are qualified to provide pharmaceutical education, monitor therapy results, contribute to decrease the risk of side events and prevent drug interactions \citep{Ansari2010,Bosse2012}. It is vital to promote initiatives which contribute to a more efficient resource allocation and a more widespread access to health care.

 Something that has been facilitating and streamlining the life of the population has been the use of communication technologies, such as of smartphone and tablets, with its usage quickly increasing in the last years. With this increase, the number and functions of health and healthcare related apps are also increasing and gaining wider adoption \citep{Bol2018}. These types of apps are also known as mHealth, an abbreviation for mobile health \citep{Adibi2015}. Some of most popular categories for these types of apps include: fitness, nutrition and medication management \citep{BinDhim2015}.

 A study by \citet{Krebs2015} conducted with 1604 mobile phone users throughout the United States, shows that more than half (58.23\%) of the surveyed users had downloaded at least one health-related app. In terms of the impact such apps can have on their users, there are apps for smoking cessation \citep{Abroms2011}, ophthalmology testing tools \citep{Chhablani2012}, for diabetes, obesity and weight management \citep{Wharton2014}, for asthma self-management \citep{Huckvale2012}, amongst many others. However, many apps are not backed by scientific data, and so can have a negative impact on their users \citep{Pandey2013}. There is a need to improve the accuracy of the information and advice given by health-related smartphone applications and encourage participation by health-care agencies to ensure patient safety.

 \citet{Wharton2014} compared the usage of a smartphone app for dietary self-monitoring against traditional diet counseling and entry methods, during an 8-week trial. Three methods were compared: the mobile app ``Lose It!'', a memo feature on a smartphone and the traditional paper-and-pencil method. The results show that the group using the weight loss app recorded dietary data more consistently that the other two groups, which shows one of the advantages of health-related apps.

  \citet{Cafazzo2012} developed an mHealth app for the self-management of type 1 diabetes adolescent patients, with a user-centered design developed through interviews made to the patients and respective family caregivers. The app allowed for simple and automated transfer of glucometer readings through a Bluetooth connection to the device to facilitate data entry. In addition, it had social and gamification features in order to increase interest and adherence. The results showed that the daily average frequency of blood glucose measurement increased by 50\%. In addition, the user satisfaction was high, with most stating they would continue to use the app. There is significant literature on the benefits of these type apps for the self-management of both type 1 and type 2 diabetes \citep{Quinn2008,Scheibe2015,Tran2012}.

 Medication nonadherence is a very common, complex, and costly problem that contributes to poor treatment outcomes and consumes many health care resources. Nonadherence greatly increases the risks of complications and comes with greatly increased health costs \citep{Briesacher2007,Ho2006,Ho2008}. Some apps assist with the management of medications and try to improve medication adherence through various methods such as alarms \citep{Dayer2013, DiDonato2015}.

 Smartphone apps also have the potential to detect some diseases and disorders, such as depression. \citet{BinDhim2015} developed an app which could detect depression through a questionnaire and which would give out personalized recommendation based on the results. Patients with a high depression score were advised by the app to take their results to their healthcare professional for further assessment. They were also provided with an explanation of the results as well as links to education websites about depression.

 Moreover, health care professionals, such as pharmacists, have also been targeted by some studies which show that such professionals can also benefit from health-related apps \citep{Davies2014}. The availability of mHealth apps targeted at pharmacists can lead to a reduction in the time taken to carry out a particular service or task, which is a key benefit to pharmacies. According to \citet{Davies2014}, to be a useful healthcare tool, a mobile app should help with the decision-making process, facilitate the pharmacist and staff, act as a communication aid and provide support and education to the patient.

 \citet{Haffey2014} reviewed smartphone apps which could help doctors with the prescriptions they make. According to the authors, junior doctors write many of the hospital prescriptions, with studies showing that doctors in their first 2 years of practice have a rate of prescription errors between 7\% and 10\%. The search which was made showed there was a large and diverse number of smartphone apps available for supporting the prescribing practice, despite the quality variance among them.

\section{Survey and Data Analysis}

 An online questionnaire was applied in which 271 responses were obtained. Table 1 shows the characterization of the respondents. The majority of them were Portuguese (96.3\%), and the age varies between 18 and 58 years old, with M=26.60 (SD=8.628). Regarding their gender, 55\% was female, and 45\% was male. 40.2\% has a college degree, 29.2\% has a master's degree, 28.4\% has a high school level, and 2.2\% has a Ph.D. Additionally, 65.7\% are students, 18.5\% are student workers, 13.7\% are workers and only 2.2\% are unemployed.

\begin{table}

\centering
\caption{Respondents characterization}
\begin{tabular}{|p{1.2in}|p{0.9in}|p{1.1in}|} \hline 

\textbf{} & \textbf{N} & \textbf{\%} \\ \hline 
Nationality & \multicolumn{2}{|p{1.9in}|}{} \\ \hline 
1     Portuguese & 261 & 96.3 \\ \hline 
     Other & 10 & 3.7 \\ \hline 
Age &  &  \\ \hline 
     M (DP) & \multicolumn{2}{|p{1.9in}|}{26.60 (8.628)} \\ \hline 
     Min-Max & \multicolumn{2}{|p{1.9in}|}{18 -- 58} \\ \hline 
Gender & \multicolumn{2}{|p{1.9in}|}{} \\ \hline 
     Female & 149 & 55.0 \\ \hline 
     Male & 122 & 45.0 \\ \hline 
School level  & \multicolumn{2}{|p{1.9in}|}{} \\ \hline 
     High school & 77 & 28.4 \\ \hline 
     College degree & 109 & 40.2 \\ \hline 
     Master's degree & 79 & 29.2 \\ \hline 
     Ph.D. & 6 & 2.2 \\ \hline 
Occupation & \multicolumn{2}{|p{1.9in}|}{} \\ \hline 
     Student & 178 & 65.7 \\ \hline 
     Student worker & 50 & 18.5 \\ \hline 
     Worker & 37 & 13.7 \\ \hline 
     Unemployed & 6 & 2.2 \\ \hline 
\end{tabular}

\end{table}

 When the workers were questioned (N=37), 70.3\% of them affirmed to be technicians, 21.6\% are managers, and 8.1\% are commercials (Table 2). These are distributed by different departments: 25.0\% in research, 25.0\% in tourism, 22.2\% in the Technical Department, 13.9\% in Healthcare, 8.3\% in Financial Management and 5.6\% in the Commercial Department.

\begin{table}

\centering
\caption{ Workers' situation}
\begin{tabular}{|p{1.2in}|p{0.9in}|p{1.1in}|} \hline 
\textbf{} & \textbf{N} & \textbf{\%} \\ \hline 
Workers' occupation & \multicolumn{2}{|p{1.9in}|}{} \\ \hline 
     Technician & 26 & 70.3 \\ \hline 
     Management & 8 & 21.6 \\ \hline 
     Commercial & 3 & 8.1 \\ \hline 
Workers' department* & \multicolumn{2}{|p{1.9in}|}{} \\ \hline 
     Research & 9 & 25.0 \\ \hline 
     Tourism & 9 & 25.0 \\ \hline 
     Technician & 8 & 22.2 \\ \hline 
     Health & 5 & 13.9 \\ \hline 
     Financial management & 3 & 8.3 \\ \hline 
     Commercial & 2 & 5.6 \\ \hline 
\end{tabular}

*N=36

\end{table}

 Regarding students (228 respondents), the majority lives in the urban area (78.1\%), and 21.9\% lives in rural area (Table 3). 74.1\% of the students are studying to obtain the master's degree, 19.7\% to obtain the Ph.D. and 6.1\% to obtain the college degree. The study area of the students is engineering (48.5\%), health (23.3\%), sciences (14.5\%), economics and management (10.6\%), and arts and architecture (3.1\%).

\begin{table}

\centering
\caption{Students' situation}
\begin{tabular}{|p{1.3in}|p{0.9in}|p{1.1in}|} \hline 
\textbf{} & \textbf{N} & \textbf{\%} \\ \hline 
Students' living area & \multicolumn{2}{|p{2.1in}|}{} \\ \hline 
     Urban & 178 & 78.1 \\ \hline 
     Rural & 50 & 21.9 \\ \hline 
Students' level objective & \multicolumn{2}{|p{2.1in}|}{} \\ \hline 
     College degree & 14 & 6.1 \\ \hline 
     Master's degree & 169 & 74.1 \\ \hline 
     Ph.D. & 45 & 19.7 \\ \hline 
Students' study area* & \multicolumn{2}{|p{2.1in}|}{} \\ \hline 
     Engineering & 110 & 48.5 \\ \hline 
     Health & 53 & 23.3 \\ \hline 
     Sciences & 33 & 14.5 \\ \hline 
     Economy and management & 24 & 10.6 \\ \hline 
     Arts and architecture & 7 & 3.1 \\ \hline 
\end{tabular}

*N=227
\end{table}

 When asked about the past usage of applications for support in the contact/location of pharmacies (results in Table 4), 63.1\% responded no, and 36.9\% (100 respondents) answered yes. The respondents who said yes, 95.0\% of them consider to search for a pharmacy-related subject less than five times per month, 1.0\% between five and ten times per month, and 4.0\% more than ten times per month. Additionally, these respondents were asked about the need for more information about in-service pharmacies or available drugs to achieve higher efficiency in obtaining their necessities. 11\% of them said never, 40.0\% rarely, 38.0\% sometimes, 10\% frequently, and 1\% always.

 Despite these opinions, 54\% of these respondents consider to know any software that allows the location of pharmacies in service: 77.8\% knows ``Farm\'{a}cias de Servi\c{c}o,'' 63\% knows ``Farm\'{a}cias Portuguesas'' and 5.6\% indicates other software as Google.

\begin{table}

\centering
\caption{Searching for pharmacies}
\begin{tabular}{|p{1.7in}|p{0.8in}|p{1.2in}|} \hline 
\textbf{} & \textbf{N} & \textbf{\%} \\ \hline 
Difficulty to find a pharmacy & \multicolumn{2}{|p{2.0in}|}{} \\ \hline 
     No & 171 & 63.1 \\ \hline 
     Yes & 100 & 36.9 \\ \hline 
Frequency of searching a pharmacy & \multicolumn{2}{|p{2.0in}|}{} \\ \hline 
     Less than 5 times per month & 95 & 95.0 \\ \hline 
     Between 5 and 10 times per month & 1 & 1.0 \\ \hline 
     More than 10 times per month & 4 & 4.0 \\ \hline 
Necessity for more information about in-service pharmacies or available drugs & \multicolumn{2}{|p{2.0in}|}{} \\ \hline 
     Never & 11 & 11.0 \\ \hline 
     Rarely & 40 & 40.0 \\ \hline 
     Sometimes & 38 & 38.0 \\ \hline 
     Frequently & 10 & 10.0 \\ \hline 
     Always & 1 & 1.0 \\ \hline 
Use of software to locate pharmacies & \multicolumn{2}{|p{2.0in}|}{} \\ \hline 
     No & 46 & 46.0 \\ \hline 
     Yes & 54 & 54.0 \\ \hline 
         Which? & \multicolumn{2}{|p{2.0in}|}{} \\ \hline 
         ``Farm\'{a}cias de servi\c{c}o'' & 42 & 77.8 \\ \hline 
         ``Farm\'{a}cias portuguesas'' & 34 & 63.0 \\ \hline 
         Other & 3 & 5.6 \\ \hline 
\end{tabular}

\end{table}

 Based on Table 5, 51.0\% of the respondents who used applications for support in the contact/location of pharmacies, find difficulties to find all the drugs in their prescription in a single pharmacy. The main difficulties mentioned are:

\begin{enumerate}
\item  sometimes the nearest or open pharmacy does not have all the necessary drugs available (60\%), 
\item  the need to order unavailable drugs and with it a considerable waiting time (25\%),
\item  it is more difficult to get drugs in weekends or holiday periods (24\%),
\item   it takes a long time to find what the respondents want, and sometimes they have to go to several pharmacies (21\%),
\item   software systems are not very explanatory of pharmacy search results or are outdated (15\%),
\item   pharmacies unavailability (1\%).
\end{enumerate}

 To solve these difficulties, 10\% has already previously contacted the pharmacy in service in order to check in advance if all the drugs in their prescription are available and do not waste time in traveling.

\begin{table}

\centering
\caption{Difficulties to find drugs}
\begin{tabular}{|p{1.9in}|p{0.7in}|p{1.1in}|} \hline 
\textbf{} & \textbf{N} & \textbf{\%} \\ \hline 
Difficulty to find drugs in a single pharmacy? & \multicolumn{2}{|p{1.8in}|}{} \\ \hline 
     No & 49 & 49.0 \\ \hline 
     Yes & 51 & 51.0 \\ \hline 
         Which? & \multicolumn{2}{|p{1.8in}|}{} \\ \hline 
sometimes the nearest or open pharmacy does not have all the necessary drugs available & 60 & 60.0 \\ \hline 
the need to order unavailable drugs and with it a considerable waiting time & 25 & 25.0 \\ \hline 
more difficult to get drugs in a weekend or holiday periods & 24 & 24.0 \\ \hline 
it takes a long time to find what the respondents want, and sometimes they have to go to several pharmacies & 21 & 21.0 \\ \hline 
software systems are not very explanatory of pharmacy search results or are outdated & 15 & 15.0 \\ \hline 
pharmacies unavailability & 1 & 1.0 \\ \hline 
Previous contact to a pharmacy? & \multicolumn{2}{|p{1.8in}|}{} \\ \hline 
     No & 90 & 90.0 \\ \hline 
     Yes & 10 & 10.0 \\ \hline 
\end{tabular}

\end{table}

 In case the respondent has to search in applications or on the internet about a particular local pharmacy (Table 6), 27.3\% consider that they only search results less or equal than 5km away, 8.1\% search results for more than 5km, 31.3\% search all results near them, 27.3\% search for 3 or fewer results, and 6.1\% search more than 3 results near them.

 Considering the scenario of the respondent going to a local pharmacy and, after arriving there, verifying that the same does not have all of their medicines available:

\begin{enumerate}
\item  59.0\% search or inquire about nearby pharmacies to complete their prescription, 

\item  32\% give up on going to a new pharmacy and order the necessary drugs, getting on the waiting list,

\item  20\% call other nearby pharmacies to see if they have the necessary drugs,

\item  16.0\% try asking someone for advice on the subject or more information.
\end{enumerate}

 Due to the difficulties currently encountered by the majority of respondents, we proposed a solution to the respondents: a system that could provide communication with pharmacies in order to submit the prescription and know in a few minutes, which closest pharmacy could provide all the drugs in the prescription. We obtained the following measures of interest:

\begin{enumerate}
\item  only 1.0\% have no interest in a new system of contact with pharmacies,

\item  6.0\% have low interest,

\item  39\% have some interest in the proposed system,

\item  moreover, 54\% of them have high interest.
\end{enumerate}

\begin{table}

\centering
\caption{ Behavior of respondents to find drugs}
\begin{tabular}{|p{2.1in}|p{0.5in}|p{1.1in}|} \hline 
\textbf{} & \textbf{N} & \textbf{\%} \\ \hline 
How to search in applications for a pharmacy?* & \multicolumn{2}{|p{1.6in}|}{} \\ \hline 
     Less or equal to 5km search & 27 & 27.3 \\ \hline 
     More than 5km search & 8 & 8.1 \\ \hline 
      Search for all near me & 31 & 31.3 \\ \hline 
     Less or equal to 3 near me & 27 & 27.3 \\ \hline 
     More than 3 near me & 6 & 6.1 \\ \hline 
Imagine you go to a local pharmacy and, after arriving there, you verify that the same does not have all of their drugs available. What do you do? & \multicolumn{2}{|p{1.6in}|}{} \\ \hline 
search or inquire about nearby pharmacies to complete the prescription & 59 & 59.0 \\ \hline 
give up on going to a new pharmacy and order the necessary drugs, getting on waiting list & 32 & 32.0 \\ \hline 
call other nearby pharmacies to see if they have the necessary drugs & 20 & 20.0 \\ \hline 
try asking someone for advice on the subject or more information & 16 & 16.0 \\ \hline 
Interest in a new system & \multicolumn{2}{|p{1.6in}|}{} \\ \hline 
     Without interest & 1 & 1.0 \\ \hline 
     Low interest & 6 & 6.0 \\ \hline 
     Some interest & 39 & 39.0 \\ \hline 
     High interest & 54 & 54.0 \\ \hline 
\end{tabular}

*N=99

\end{table}

 After the descriptive analysis of the sample, it is essential to perceive the influence of some factors in the opinion of the respondents. Thus, a cross-tabulation analysis of the information was performed using the chi-square test, whose null hypothesis is that the variables are independent. Additionally, the chi-square test provides two statistics that indicate whether the variables are associated or independent: a chi-square statistic and a p-value.

 Thus, if the p-value is less than or equal to 0.05 (level of significance considered as acceptable), the null hypothesis is rejected, and the variables are associated. If the p-value is more significant than 0.05, the null hypothesis is not rejected, and it is possible to conclude that variables are independent.

 Table 7 presents the main results of the influence of the age in some opinions of the respondents. Regarding the first comparison (age vs. difficulty to obtain drugs), the null hypothesis is rejected, and variables are associated. This means that young people have fewer difficulties to find drugs, and older people have more difficulties, which could be explained by the fact that old people have more health problems or have already lived more experiences.

 However, it is clear that age does not influence the interest in a new system, the necessity of more information and the use of location software.

\begin{table}

\centering
\caption{ Influence of age in the respondent opinion}
\begin{tabular}{|p{1.1in}|p{0.7in}|p{0.8in}|p{0.6in}|p{0.4in}|p{0.4in}|} \hline 
\multicolumn{2}{|p{1in}|}{} & \multicolumn{2}{|p{1.5in}|}{Age} & X${}^{2}$ & p \\ \hline 
\multicolumn{2}{|p{1in}|}{} & Less or equal to 23 & More than 23 &  &  \\ \hline 
Difficulty in obtaining drugs & Yes & 19 & 32 & 6.763 & \textbf{0.009} \\ \hline 
 & No & 31 & 18 &  & \textbf{} \\ \hline 
Interest in a new system & Some interest & 27 & 19 & 2.576 & 0.108 \\ \hline 
 & High interest & 23 & 31 &  &  \\ \hline 
Necessity of more information & Rarely & 30 & 21 & 3.241 & 0.072 \\ \hline 
 & Frequently & 20 & 29 &  &  \\ \hline 
Use of location software & Yes & 24 & 30 & 1.449 & 0.229 \\ \hline 
 & No & 26 & 20 &  &  \\ \hline 
\end{tabular}
\end{table}

 The comparison between the necessity of more information and the difficulty in obtaining drugs (Table 8) show that these variables are associated (p=0.045$\mathrm{<}$0.05). Thus, people who had more difficulties to get drugs are precisely those who feel more necessity to have more information available. By the opposite, people with few difficulties to obtain their drugs are who rarely feel the necessity to have more information available.

\begin{table}

\centering
\caption{Relation between necessity of more information and difficulty to obtain drugs}
\begin{tabular}{|p{1.1in}|p{0.7in}|p{0.8in}|p{0.6in}|p{0.4in}|p{0.4in}|} \hline 
\multicolumn{2}{|p{1in}|}{} & \multicolumn{2}{|p{1.5in}|}{Necessity of more information} & X${}^{2}$ & p \\ \hline 
\multicolumn{2}{|p{1in}|}{} & Rarely & Frequently &  &  \\ \hline 
Difficulty in obtaining drugs & Yes & 21 & 30 & 4.019 & \textbf{0.045} \\ \hline 
 & No & 30 & 19 &  & \textbf{} \\ \hline 
\end{tabular}
\end{table}

 Table 9 present some factor that could influence the interest in a new system. Regarding the difficulty to obtain drugs, it is possible to verify that people who had to feel that, are the most interested in a new system, and people with fewer difficulties does not feel the same interest. 

 Additionally, people who had already a waste of time to find the necessary drugs have more interest in a new system. If they do not waste time to find drugs, they have less interest in a new system.

 Regarding the situation ``sometimes the open pharmacy does not have the needed drugs'' vs. ``interest in a new system,'' the null hypothesis is rejected and, then, both variables are associated. Similar to previous analysis, people who consider that sometimes the open pharmacy does not have the needed drugs have a high interest in a new system.

 Finally, people that search or inform of nearby pharmacies are the most interested in a new system (p=0.004), and the opposite also happens.

 To conclude, it turns out that the most active people in the search for pharmacies, drugs or even information are those who are most interested in a new system that easily makes all this information available.

 \begin{table}

\centering
\caption{Factors that influence the interest in a new system}

\begin{tabular}{|p{1.1in}|p{0.7in}|p{0.8in}|p{0.6in}|p{0.4in}|p{0.4in}|} \hline 
\multicolumn{2}{|p{1in}|}{} & \multicolumn{2}{|p{1.5in}|}{Interest in a new system} & X${}^{2}$ & p \\ \hline 
\multicolumn{2}{|p{1in}|}{} & Some interest & High interest &  &  \\ \hline 
Difficulty to obtain drugs & Yes & 15 & 36 & 11.530 & \textbf{0.001} \\ \hline 
 & No & 31 & 18 &  & \textbf{} \\ \hline 
Waste of time to find the necessary drugs & Yes & 5 & 16 & 5.270 & \textbf{0.022} \\ \hline 
 & No & 41 & 38 &  & \textbf{} \\ \hline 
Sometimes the open pharmacy does not have the needed drugs  & Yes & 21 & 39 & 7.307 & \textbf{0.007} \\ \hline 
 & No & 25 & 15 &  & \textbf{} \\ \hline 
I search or inform myself of nearby pharmacies & Yes & 20 & 39 & 8.484 & \textbf{0.004} \\ \hline 
 & No & 26 & 15 &  & \textbf{} \\ \hline 
\end{tabular}
\end{table}
\section{Proposed Solution}

 During a course of I.T. Engineering Degree at Faculty of Engineering, University of Porto, the students, oriented by the course teachers and the authors of the idealization of such system, proposed Medloc, a new software system. This new system would support both the patients and the pharmacists in achieving a better and more efficient communication.

 Medloc was idealized to be an app that will allow:

\begin{enumerate}
\item  Fast and efficient communication between users and pharmacies

\item  Better management done by the users of their own prescription

\item  It can replace traditional face-to-face or telephone answering service, useful in periods of minimal services avoiding long queues and less time spent by pharmacists over the phone which can be used to attend to patients or other important tasks 

\item  Information on the availability of the medicines required by the patients on his smartphone or computer instead of wasting time calling over the phone or taking time to search in several pharmacies

\item  Information about the location of the pharmacy
\end{enumerate}

\subsection{Prescription management}

 \textbf{}

 In Figure 1, it is possible to check that the user can insert his prescription. Additionally, the user is helped by an automatic completion of the medicines, i.e., the application has autocompleted features. The user can cancel, edit or ask for availability of prescription medicines, in corresponding buttons.  

\begin{figure}
    \centering
    \includegraphics[scale=1]{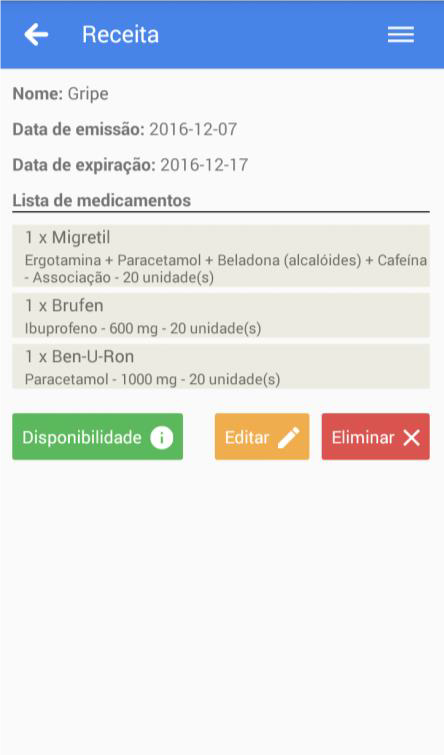}
    \caption{Prescription insertion}
    \label{fig:1}
\end{figure}

\subsection{Availability Requests}

 After pressing the availability button in the app (see the previous figure), the systems sends the request from the user app to the server, and finally, the request is forward to registered pharmacies near the user location. Figure 2 shows the list of pharmacies that received the request with the option for the user to press the button and see each pharmacy in the map. In Figure 3, it is also possible to view the app feature related to notifications, these notifications are related to responses from individual pharmacies with the availability of the prescription medicines.

\begin{figure}
    \centering
    \includegraphics[scale=1]{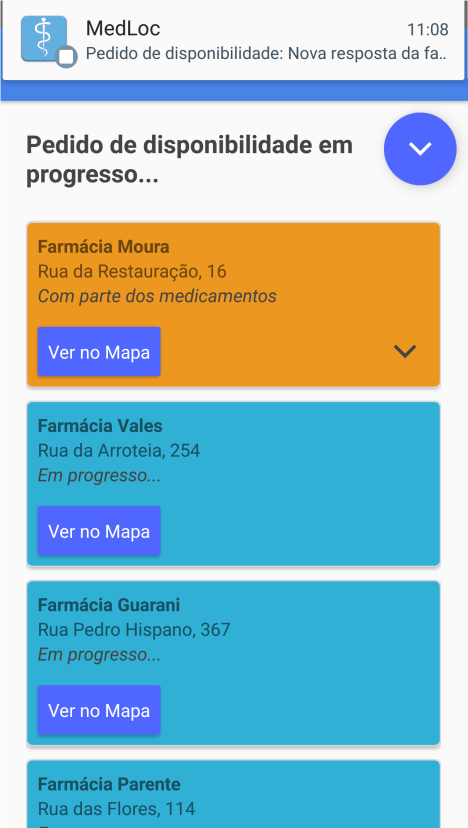}
    \caption{Availability requests}
    \label{fig:2}
\end{figure}

\subsection{ Response to drug availability requests}

 \textbf{}

 In case of no response, it is possible to check in the following figure that all pharmacies tried did not respond positively. What might happen in this case, is that the system will update its radius of search and will expand the area of search and increase the number of enquired pharmacies, after a specified period of time (for example, 10 minutes). Please see Figure 3.

\begin{figure}
    \centering
    \includegraphics[scale=1]{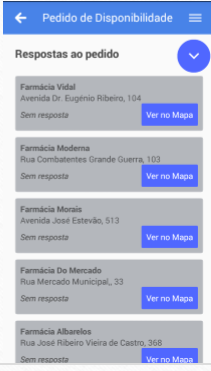}
    \caption{Availability requests response}
    \label{fig:3}
\end{figure}

\subsection{From the point of view of the Pharmacies}

 Figure 4, demonstrates a possible layout for the pharmacist application. It is visible that the pharmacist is preparing a response to the patient and user of the mobile app. In this case, the pharmacist does not have all medicines available, and he checks in the checkbox only the ones he has immediately available. It is also possible to check that the Pharmacist has one green and a red button that allows responding rapidly to requests, respectively with the green button when the pharmacist has all medicines and the red button when the pharmacist has none of the medicines in the prescription.

\begin{figure}
    \centering
    \includegraphics[scale=0.85]{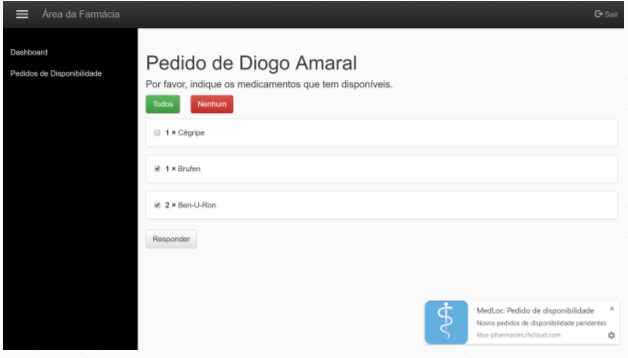}
    \caption{Pharmacy point of view and response screen}
    \label{fig:4}
\end{figure}

\subsection{Pharmacies Location}

 Please check Figure 5. In this screen, the patient can see all the pharmacies that are near him. It also shows some more information about the pharmacies.\textbf{}

\begin{figure}
    \centering
    \includegraphics[scale=1]{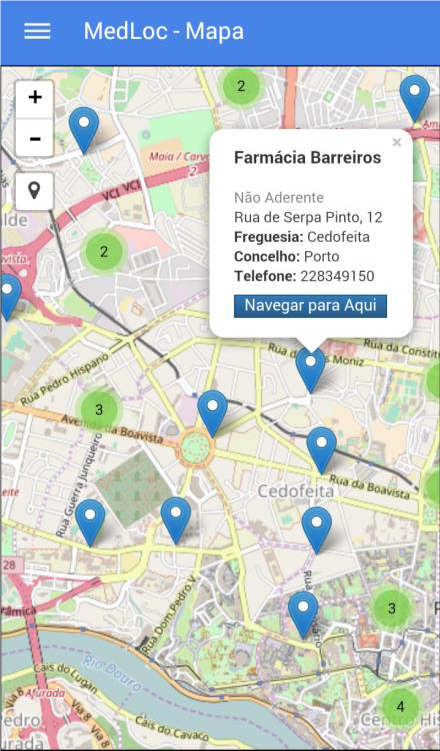}
    \caption{Map of pharmacies location}
    \label{fig:5}
\end{figure}

\section{Issues, Controversies, Problems}

 The major issues Medloc can suppress are:

\begin{enumerate}
\item  Communication between users and pharmacies only on-site or by telephone

\item  Inexistence of information on the availability of medicines at nearby pharmacies

\item  Long queues in periods of minimum services or weekends

\item  App will be useful for users who are deaf and/or mute
\end{enumerate}

 The major problems Medloc can have against are:

\begin{enumerate}
\item  Adoption of a new system may involve some adjustments of procedures by pharmacies

\item  The elderly population is not proficient with apps used in smartphones or even computers 
\end{enumerate}

\section{Conclusion}

 Concluding, in this document we presented a novel view for the needs of patients and pharmacists. We demonstrated that the general opinion of the public leans towards a new needed system to provide communication between both sides of the chain.

 Also in this document, we proposed a new system, MEDLOC, simpler and faster communication between patient/user and pharmacy. We concluded that this software provided means to find the nearest pharmacies to a certain location, with better management of the prescriptions. Thus, with the possibility to question (online) the pharmacy about medicines available in stock, adhering pharmacies are notified of new availability requests and can respond to those requests. The user is notified through both the site and the mobile app when new responses to their requests appear in an efficient and transparent manner. Such a system can avoid the loss of commuting time for patients and also service time for pharmacists. 

 Future work would be to test pilot the MEDLOC prototype in a working environment in a chain of pharmacies or agreement with a country's Pharmacy supervising institution. Further work might also be achieved regarding the possibility for the patient to make reservation of medicine(s) or even request the delivery of medicines to their house. Considering the possibility of major retail enterprises being able to provide medicines delivery, this could be an essential step to pharmacies retail system and improvement of their business.

\section*{Acknowledgments}
The authors would like to thank FEUP (Faculty of Engineering, University of Porto) for the opportunity to propose and be granted the development of the mentioned prototype. Thus, since the development was in the context of a course of the Master's in Informatics Engineering, we would like to thank to the Professors involved, Professor Ademar Aguiar and Professor Rui Maranh\~{a}o. Rui Portocarrero Sarmento directs a special acknowledgment to all the students involved in the development of the prototype, Pedro C\^{a}mara, Ant\'{o}nio Casimiro, Rui Cardoso, Pedro Silva, Catarina Amaral, and Diogo Amaral. Rui Portocarrero Sarmento also gratefully acknowledges funding from FCT (Portuguese Foundation for Science and Technology) through a Ph.D. grant (SFRH/BD/119108/2016). The authors want to thank also to the reviewers for the constructive reviews provided in the development of this publication.

\bibliographystyle{apalike}
\bibliography{Report}

\begin{thebibliography}{}

\bibitem[Abroms et~al., 2011]{Abroms2011}
Abroms, L.~C., Padmanabhan, N., Thaweethai, L., and Phillips, T. (2011).
\newblock {iPhone Apps for Smoking Cessation}.
\newblock {\em American Journal of Preventive Medicine}, 40(3):279--285.

\bibitem[Adibi, 2015]{Adibi2015}
Adibi, S. (2015).
\newblock {\em {Mobile health: a technology road map}}, volume~5.
\newblock Springer.

\bibitem[Ansari, 2010]{Ansari2010}
Ansari, J. (2010).
\newblock {Drug interaction and pharmacist.}
\newblock {\em Journal of young pharmacists : JYP}, 2(3):326--31.

\bibitem[BinDhim et~al., 2015]{BinDhim2015}
BinDhim, N.~F., Shaman, A.~M., Trevena, L., Basyouni, M.~H., Pont, L.~G., and
  Alhawassi, T.~M. (2015).
\newblock {Depression screening via a smartphone app: cross-country user
  characteristics and feasibility.}
\newblock {\em Journal of the American Medical Informatics Association :
  JAMIA}, 22(1):29--34.

\bibitem[Bol et~al., 2018]{Bol2018}
Bol, N., Helberger, N., and Weert, J. C.~M. (2018).
\newblock {Differences in mobile health app use: A source of new digital
  inequalities?}
\newblock {\em The Information Society}, 34(3):183--193.

\bibitem[Bosse et~al., 2012]{Bosse2012}
Bosse, N., Machado, M., and Mistry, A. (2012).
\newblock {Efficacy of an over-the-counter intervention follow-up program in
  community pharmacies}.
\newblock {\em Journal of the American Pharmacists Association},
  52(4):535--540.

\bibitem[Briesacher et~al., 2007]{Briesacher2007}
Briesacher, B.~A., Gurwitz, J.~H., and Soumerai, S.~B. (2007).
\newblock {Patients At-Risk for Cost-Related Medication Nonadherence: A Review
  of the Literature}.
\newblock {\em Journal of General Internal Medicine}, 22(6):864--871.

\bibitem[Cafazzo et~al., 2012]{Cafazzo2012}
Cafazzo, J.~A., Casselman, M., Hamming, N., Katzman, D.~K., and Palmert, M.~R.
  (2012).
\newblock {Design of an mHealth app for the self-management of adolescent type
  1 diabetes: a pilot study.}
\newblock {\em Journal of medical Internet research}, 14(3):e70.

\bibitem[Chhablani et~al., 2012]{Chhablani2012}
Chhablani, J., Kaja, S., and Shah, V.~A. (2012).
\newblock {Smartphones in ophthalmology.}
\newblock {\em Indian journal of ophthalmology}, 60(2):127--31.

\bibitem[Davies et~al., 2014]{Davies2014}
Davies, M.~J., Collings, M., Fletcher, W., and Mujtaba, H. (2014).
\newblock {Pharmacy Apps: a new frontier on the digital landscape?}
\newblock {\em Pharmacy practice}, 12(3):453.

\bibitem[Dayer et~al., 2013]{Dayer2013}
Dayer, L., Heldenbrand, S., Anderson, P., Gubbins, P.~O., and Martin, B.~C.
  (2013).
\newblock {Smartphone medication adherence apps: Potential benefits to patients
  and providers}.
\newblock {\em Journal of the American Pharmacists Association},
  53(2):172--181.

\bibitem[DiDonato et~al., 2015]{DiDonato2015}
DiDonato, K.~L., Liu, Y., Lindsey, C.~C., Hartwig, D.~M., and Stoner, S.~C.
  (2015).
\newblock {Community pharmacy patient perceptions of a pharmacy-initiated
  mobile technology app to improve adherence}.
\newblock {\em International Journal of Pharmacy Practice}, 23(5):309--319.

\bibitem[Haffey et~al., 2014]{Haffey2014}
Haffey, F., Brady, R. R.~W., and Maxwell, S. (2014).
\newblock {Smartphone apps to support hospital prescribing and pharmacology
  education: a review of current provision}.
\newblock {\em British Journal of Clinical Pharmacology}, 77(1):31--38.

\bibitem[Ho et~al., 2008]{Ho2008}
Ho, P.~M., Magid, D.~J., Shetterly, S.~M., Olson, K.~L., Maddox, T.~M.,
  Peterson, P.~N., Masoudi, F.~A., and Rumsfeld, J.~S. (2008).
\newblock {Medication nonadherence is associated with a broad range of adverse
  outcomes in patients with coronary artery disease}.
\newblock {\em American Heart Journal}, 155(4):772--779.

\bibitem[Ho et~al., 2006]{Ho2006}
Ho, P.~M., Rumsfeld, J.~S., Masoudi, F.~A., McClure, D.~L., Plomondon, M.~E.,
  Steiner, J.~F., and Magid, D.~J. (2006).
\newblock {Effect of Medication Nonadherence on Hospitalization and Mortality
  Among Patients With Diabetes Mellitus}.
\newblock {\em Archives of Internal Medicine}, 166(17):1836.

\bibitem[Huckvale et~al., 2012]{Huckvale2012}
Huckvale, K., Car, M., Morrison, C., and Car, J. (2012).
\newblock {Apps for asthma self-management: a systematic assessment of content
  and tools.}
\newblock {\em BMC medicine}, 10:144.

\bibitem[Krebs and Duncan, 2015]{Krebs2015}
Krebs, P. and Duncan, D.~T. (2015).
\newblock {Health App Use Among US Mobile Phone Owners: A National Survey.}
\newblock {\em JMIR mHealth and uHealth}, 3(4):e101.

\bibitem[Pandey et~al., 2013]{Pandey2013}
Pandey, A., Hasan, S., Dubey, D., and Sarangi, S. (2013).
\newblock {Smartphone Apps as a Source of Cancer Information: Changing Trends
  in Health Information-Seeking Behavior}.
\newblock {\em Journal of Cancer Education}, 28(1):138--142.

\bibitem[Quinn et~al., 2008]{Quinn2008}
Quinn, C.~C., Clough, S.~S., Minor, J.~M., Lender, D., Okafor, M.~C., and
  Gruber-Baldini, A. (2008).
\newblock {WellDoc {\textless}sup{\textgreater}™{\textless}/sup{\textgreater}
  Mobile Diabetes Management Randomized Controlled Trial: Change in Clinical
  and Behavioral Outcomes and Patient and Physician Satisfaction}.
\newblock {\em Diabetes Technology {\&} Therapeutics}, 10(3):160--168.

\bibitem[Scheibe et~al., 2015]{Scheibe2015}
Scheibe, M., Reichelt, J., Bellmann, M., and Kirch, W. (2015).
\newblock {Acceptance factors of mobile apps for diabetes by patients aged 50
  or older: a qualitative study.}
\newblock {\em Medicine 2.0}, 4(1):e1.

\bibitem[Tran et~al., 2012]{Tran2012}
Tran, J., Tran, R., and White, J.~R. (2012).
\newblock {Smartphone-based glucose monitors and applications in the management
  of diabetes: An overview of 10 salient "apps" and a novel
  smartphone-connected blood glucose monitor}.
\newblock {\em Clinical Diabetes}, 30(4):173--178.

\bibitem[Viswanathan et~al., 2015]{Viswanathan2015}
Viswanathan, M., Kahwati, L.~C., Golin, C.~E., Blalock, S.~J., Coker-Schwimmer,
  E., Posey, R., and Lohr, K.~N. (2015).
\newblock {Medication Therapy Management Interventions in Outpatient Settings}.
\newblock {\em JAMA Internal Medicine}, 175(1):76.

\bibitem[Wharton et~al., 2014]{Wharton2014}
Wharton, C.~M., Johnston, C.~S., Cunningham, B.~K., and Sterner, D. (2014).
\newblock {Dietary Self-Monitoring, But Not Dietary Quality, Improves With Use
  of Smartphone App Technology in an 8-Week Weight Loss Trial}.
\newblock {\em Journal of Nutrition Education and Behavior}, 46(5):440--444.

\end{thebibliography}

\end{document}